\documentclass[twocolumn]{revtex4}

\input epsf
\usepackage{amsmath,amssymb}
\usepackage{graphicx}

\draft

\begin{document}
\titlepage
\title{Palatini formulation of L(R) gravity}
\author{Peng Wang$^1$ \footnote{E-mail: pewang@eyou.com} and Xin-He Meng$^{1,2}$ \footnote{E-mail: xhm@physics.arizona.edu\\
xhmeng@phys.nankai.edu.cn}
} \affiliation{1.Department of Physics, Nankai University, Tianjin
300071, P.R.China\\ 2. Department of
Physics, University of Arizona, Tucson, AZ 85721 USA}

\begin{abstract}
We review the Palatini formulation of the higher-derivative
gravity of the $L(R)$ form and its applications in cosmology.
\end{abstract}

\maketitle

\section{Introduction}
It now seems well-established that the expansion of our universe
is currently in an accelerating phase. The most direct evidence
for this is from the measurements of type Ia supernova
\cite{Perlmutter}. Other indirect evidences such as the
observations of CMB by the WMAP satellite \cite{Spergel},
large-scale galaxy surveys by 2dF and SDSS \cite{hui} also seem
supporting this.

But now the mechanisms responsible for this acceleration are not
very clear. Many authors introduce a mysterious cosmic fluid
called dark energy to explain this (see Ref.\cite{Peebles} for a
review). On the other hand, some authors suggest that maybe there
does not exist such mysterious dark energy, but instead  the observed
cosmic acceleration is a signal of our first real lack of
understanding of gravitational physics \cite{Lue}. An example is
the braneworld theory of Dvali et al. \cite{Dvali}. Recently,
there are active discussions in this direction by extending the General Relativity theory, i.e., modifying the
action for gravity either by introducing extra-dimension with brane(s) or adding correction term(s) in
view of quantum anomaly or phenomenology considerations [7-17]. Specifically, a relatively simple $1/R$ term is suggested
to be added to the action \cite{Carroll}: the so called $1/R$
gravity. It is interesting that such term may be predicted by a
string/M-theory \cite{Odintsov-string}. In Ref.\cite{Vollick},
Vollick used Palatini variational principle to derive the field
equations for $1/R$ gravity. In Ref.\cite{Dolgov}, Dolgov \emph{et
al.} argued that the fourth order field equations following from
the metric variation suffer serious instability problem. If this
is indeed the case, the Palatini formulation appears quite
appealing, because the second order field equations following from
Palatini variation are free of this sort of instability
\cite{Wang}. However, it is also interesting to note that quantum
effects may resolve the instabilities, see
Ref.\cite{odintsov-insta}. In this paper, we will review deriving
the full Modified Friedmann equation for Palatini formulation of
the general $L(R)$ type gravity. Then we will discuss the applications of this
derivation to $1/R$, $R^2$, $1/R+R^2$ and $\ln R$ gravity models and their cosmological implyings.
Recently, Palatini formulation of $L(R)$ gravity was also
considered in Ref.\cite{new}.

\section{The Modified Friedmann Equation}
In general, when handled in Palatini formulation, one considers
the action to be a functional of the metric $\bar{g}_{\mu\nu}$ and
a connection $\hat{\bigtriangledown}_{\mu}$ which is another
independent variable besides the metric. The resulting modified
gravity action can be written as
\begin{equation}
S[\bar{g}_{\mu\nu}, \hat{\bigtriangledown}_{\mu}]=\int
d^4x\sqrt{-\bar{g}}\frac{1}{2\kappa^2}L(\hat R)+S_m\label{1}
\end{equation}
where we use the metric signature $\{-,+,+,+\}$, $\kappa^2=8\pi
G$, $\hat{R}_{\mu\nu}$ is the Ricci tensor of the connection
$\hat{\bigtriangledown}_{\mu}$,
$\hat{R}=\bar{g}^{\mu\nu}\hat{R}_{\mu\nu}$, and $S_m$ is the
matter action.

Varying the action (1) with respect to $\bar g_{\mu\nu}$ gives
\begin{equation}
L'(\hat R)\hat R_{\mu\nu}-\frac{1}{2}L(\hat R)\bar
g_{\mu\nu}=\kappa^2 T_{\mu\nu}\label{2.2}
\end{equation}
where a prime denotes differentiation with respect to $\hat R$ and
$T_{\mu\nu}$ is the energy-momentum tensor given by
\begin{equation}
T_{\mu\nu}=-\frac{2}{\sqrt{-\bar g}}\frac{\delta S_M}{\delta \bar
g^{\mu\nu}}\label{2.3}
\end{equation}
where $S_M$ is the matter action. For a universe filled with
perfect fluid, $T^{\mu}_{\nu}=\{-\rho,p,p,p\}$. Note that the
local conservation of energy momentum $\bar\nabla_\mu
T^{\mu\nu}=0$ is a result of the covariance of the action (1)
 and Neother theorem, thus it is independent of
the gravitational field equations. Then the energy conservation
equation $\dot\rho+3H(\rho+p)=0$ is unchanged.

In the Palatini formulation, the connection is associated with
$\hat g_{\mu\nu}\equiv L'(\hat R)\bar g_{\mu\nu}$, which is known
from varying the action with respect to $\hat \Gamma
^{\lambda}_{\mu\nu}$. Thus the Christoffel symbol with respect to
$\hat g_{\mu\nu}$ is given in terms of the Christoffel symbol with
respect to $\bar g_{\mu\nu}$ by
\begin{equation}
\hat\Gamma ^{\lambda}_{\mu\nu}=\bar\Gamma
^{\lambda}_{\mu\nu}+\frac{1}{2L'}[2\delta
^{\lambda}_{(\mu}\partial _{\nu)}L'-\bar g_{\mu\nu}\bar
g^{\lambda\sigma}\partial _{\sigma}L']\label{Christoffel}
\end{equation}

Thus the Ricci curvature tensor is given by
\begin{eqnarray}
\hat R_{\mu\nu}=\bar R_{\mu\nu}+\frac{3}{2}(L')^{-2}\bar\nabla
_{\mu}L'\bar\nabla _{\nu}L' -(L')^{-1}\bar\nabla _{\mu}\bar\nabla
_{\nu}L'\\\nonumber-\frac{1}{2}(L')^{-1}\bar g_{\mu\nu}\bar\nabla
_{\sigma}\bar\nabla ^{\sigma}L'\label{Ricci}
\end{eqnarray}
and
\begin{equation}
\hat R=\bar R-3(L')^{-1}\bar\nabla _{\mu}\bar\nabla ^{\mu}
L'+\frac{3}{2}(L')^{-2}\bar\nabla_{\mu}L'\bar\nabla^{\mu}L'\label{scalar}
\end{equation}
where $\bar R_{\mu\nu}$ is the Ricci tensor with respect to $\bar
g_{\mu\nu}$ and $\hat R=\bar g^{\mu\nu}\hat R_{\mu\nu}$. Note by
contracting Eq.(\ref{2.2}), we get:
\begin{equation}
L'(\hat R)\hat R-2L(\hat R)=\kappa^2 T\label{R(T)}
\end{equation}
Assume we can solve $\hat R$ as a function of $T$ from
Eq.(\ref{R(T)}). Thus Eqs.(\ref{Ricci}), (\ref{scalar}) do define
the Ricci tensor with respect to $h_{\mu\nu}$.

We will consider the general Robertson-Walker metric (Note that
this is an ansatz for $\bar g_{\mu\nu}$ and is the result of the
assumed homogenization and isotropy of the universe, thus its form
is independent of the gravity theory):
\begin{equation}
ds^2=-dt^2+a(t)^2(\frac{dr^2}{1-kr^2}+r^2(d\theta^2+\sin^2\theta
d\phi^2))\label{metric}
\end{equation}
where $k$ is the spatial curvature and $k=-1,0,1$ correspond to
open, flat and closed universe respectively. The $a(t)$ is called
the scale factor of the universe.

From equations (\ref{metric}), and (\ref{Ricci}), we can get the
non-vanishing components of the Ricci tensor:
\begin{equation}
\hat
R_{00}=-3\frac{\ddot{a}}{a}+\frac{3}{2}(L')^{-2}(\partial_0{L'})^2-\frac{3}{2}(L')^{-1}\bar
\nabla_0\bar\nabla_0L'\label{R00}
\end{equation}
\begin{eqnarray}
\hat R_{ij}=[a\ddot{a}+2\dot{a}^2+2k+(L')^{-1}\bar\Gamma
^0_{ij}\partial_0L' \\\nonumber+\frac{a^2}{2}(L')^{-1}\bar
\nabla_0\bar\nabla_0L']\delta_{ij}\label{ij}
\end{eqnarray}
where a dot denotes differentiation with respect to $t$.

Substituting equations (\ref{R00}) and (\ref{ij}) into the field
equations (\ref{2.2}), we can get
\begin{eqnarray}
6H^2+3H(L')^{-1}\partial_0L'+\frac{3}{2}(L')^{-2}(\partial_0L')^2+6\frac{k}{a^2}\\\nonumber=\frac{\kappa^2
(\rho+3p)+L}{L'}\label{MF}
\end{eqnarray}
where $H\equiv \dot{a}/a$ is the Hubble parameter, $\rho$ and $p$
are the total energy density and total pressure respectively.

Using the energy conservation equation $\dot\rho+3H(\rho+p)=0$, we
have
\begin{equation}
\partial_0L'=\frac{2\kappa^2}{\beta}(1-3c_s^2)(\rho+p)\label{21}
\end{equation}
where $c_s^2=dp/d\rho$ is the sound velocity.

Substituting Eq.(\ref{21}) into Eq.(\ref{MF}) we can get the
Modified Friedmann (MF) equation of $L(R)$ gravity in Palatini
formulation.

\section{Applications to Cosmology}
In this section, we review the applications of the general Modified
Friedmann equation in four specific cases to cosmology.

\subsection{$1/R$ gravity}

The Lagrangian is given by $L(\hat R)=\hat R-\mu^4/\hat R$
\cite{Carroll}. It is interesting to note that, in
Ref.\cite{Odintsov-string}, Nojiri and Odintsov have shown that
this action can be derived from string/M theory.

The MF equation follows from eq.(\ref{MF}) \cite{Wang}

\begin{equation}
H^2=\frac{\kappa \rho-\alpha(G(\frac{\kappa
\rho}{\alpha})-\frac{1}{3G(\frac{\kappa
\rho}{\alpha})})}{(1+\frac{1}{3G(\frac{\kappa
\rho}{\alpha})^2})(6+3F(\frac{\kappa
\rho}{\alpha})[1+\frac{1}{2}F(\frac{\kappa
\rho}{\alpha})])}\label{MF}
\end{equation}
where the two functions $G$ and $F$ are defined as
\begin{equation}
G(x)=-(\frac{1}{2}x+\sqrt{1+\frac{1}{4}x^2})\label{G}
\end{equation}
\begin{equation}
F(x)=\frac{x}{(G(x)^2+\frac{1}{3})\sqrt{1+\frac{1}{4}x^2}}\label{F}
\end{equation}

In Ref.\cite{Wang}, we have shown that the above MF equation can
fit the current SN Ia data at an acceptable level. However, the
effective equation of state it gives shows some pathological
behaviors.

\subsection{$R^2$ gravity}

It is a well-known result that a $R^2$ term in the Lagrangian can
drive an early universe inflation without inflaton \cite{Sta} (See
also Ref.\cite{Odintsov-book} for a comprehensive discussion of
$R^2$ gravity).

In the Palatini formulation, the MF equation is \cite{Wang2}
\begin{equation}
H^2=\frac{2\kappa^2(\rho_m+\rho_r)+\frac{(\kappa^2\rho_m)^2}{3\beta}}{(1+\frac{2\kappa^2\rho_m}{3\beta})(6+3F_0(
\frac{\kappa^2\rho_m}{\beta})(1+\frac{1}{2}F_0(\frac{\kappa^2\rho_m}{\beta}))}\label{R2MF}
\end{equation}
where the function $F_0$ is given by
\begin{equation}
F_0(x)=-\frac{2x}{1+\frac{2}{3}x}\label{F0}
\end{equation}
It is interesting to see from Eq.(\ref{R2MF}) that all the effects
of the $R^2$ term are determined by $\rho_m$. If $\rho_m=0$,
Eq.(\ref{R2MF}) simply reduces to the standard Friedmann equation.

At late cosmological times when $\kappa^2\rho_m/\beta \ll 1$,
$F_0\sim 0$, the MF equation (\ref{R2MF}) reduces to the standard
Friedmann equation:
\begin{equation}
H^2=\frac{\kappa^2}{3}(\rho_m+\rho_r)\label{}
\end{equation}
Thus from the BBN constraints on the Friedmann equation
\cite{Carroll4}, $\beta$ should be sufficiently large so that the
condition $\kappa^2\rho_m/\beta \ll 1$ is satisfied in the era of
BBN. In typical model of $R^2$ inflation, $\beta$ is often taken
to be the order of the Plank scale \cite{Sta}.

Then we will see that whatever we assume on the value of $\rho_m$
during inflation, a $R^2$ driven inflation can not happen.

Firstly, since inflation happens in very early universe, where the
temperature is typical of the $10^{15}$ Gev order, if we assume
that almost all the matter in the universe at that time is
relativistic so that $\kappa^2\rho_m/\beta \ll 1$, then as we have
shown above, the MF equation reduces to the standard Friedmann
equation and thus no inflation happens. Note that at the inflation
energy scale, all the standard model particles will be
relativistic.

Secondly, if there are enough exotic objects other than the
standard model particles that will be non-relativistic during the
inflation era so that $\kappa^2\rho_m/\beta \gg 1$. Those objects
may be primordial black holes, various topological defects which
we will not specify here. In this case, from Eq.(\ref{F0}), the MF
equation (\ref{R2MF}) will reduce to
\begin{equation}
H^2=\frac{\kappa^2\rho_m}{21}+\frac{2\beta\rho_r}{7\rho_m}+\frac{2\beta}{7}\label{}
\end{equation}
Then we can see that if the $\beta$ term dominates over the other
two terms, it will drive an inflation. But this equation is
derived under the assumption that $\beta\ll \kappa^2\rho_m$. Thus
no inflation will appear, too.

Thus, the difference between Palatini and metric formulation of
the same higher derivative gravity theory is quite obvious here.

\subsection{$1/R+R^2$ gravity}

In Ref.\cite{Odintsov-R2}, Nojiri and Odinstov showed that a
combination of the $1/R$ and $R^2$ terms can drive both the
current acceleration and inflation. The Palatini form of this
theory is studied in Ref.\cite{Wang2}.

The MF equation reads,
\begin{eqnarray}
&&H^2=\cr&&\frac{\kappa^2\rho_m+2\kappa^2\rho_r+\alpha[G(x)-\frac{1}{3G(x)}
+\frac{\alpha}{3\beta} G(x)^2]}{[1+\frac{1}{3G(x)^2}+
\frac{2\alpha}{3\beta} G(x)][6+3F(x)(1+\frac{1}{2}F(x))]}
\label{s1}
\end{eqnarray}
where $x\equiv\frac{\kappa^2\rho_m}{\alpha}$ and the two functions
$G$ and $F$ are given by
\begin{equation}
G(x)=\frac{1}{2}[x+2\sqrt{1+\frac{1}{4}x^2}]\label{s2}
\end{equation}
\begin{equation}
F(x)=\frac{(1-\frac{\alpha}{\beta}G(x)^3)x}{(G(x)^2+\frac{2\alpha}{3\beta}G(x)^3+\frac{1}{3})\sqrt{1+\frac{1}{4}x^2}}
\label{s3}
\end{equation}

In order to be consistent with observations, we should have
$\alpha\ll\beta$. We can see this in two different ways.

Firstly, when $\kappa^2\rho_m\gg\alpha$, from Eq.(\ref{s2}),
$G\sim\kappa^2\rho_m/\alpha$. From the BBN constraints, we know
the MF equation should reduce to the standard one in the BBN era
\cite{Carroll4}. This can be achieved only when $F\sim0$ and from
Eq.(\ref{s3}), this can be achieved only when $\alpha\ll\beta$ and
$1\ll\kappa^2\rho_m/\alpha\ll(\beta/\alpha)^{1/3}$.

Secondly, when $\kappa^2\rho_m\ll \alpha$, we can expand the RHS
of Eq.(\ref{s1}) to first order in $\kappa^2\rho_m/\alpha$:
\begin{equation}
H^2=\frac{\frac{11-\alpha/\beta}{8-4\alpha/\beta}\kappa^2\rho_m+\frac{3}{2-\alpha/\beta}\kappa^2\rho_r+\frac{1}{2}\alpha}
{6+\frac{9}{4-2\alpha/\beta}(1+\alpha/\beta)\frac{\kappa^2\rho_m}{\alpha}}\label{1stfull}
\end{equation}

When $\alpha\ll\beta$, this will reduces exactly to the first
order MF equation in the 1/R theory \cite{Wang}. Since we have
shown there that the MF equation in 1/R theory can fit the SN Ia
data at an acceptable level, the above MF equation can not deviate
from it too large, thus the condition $\alpha\ll\beta$ should be
satisfied.

\subsection{$\ln R$ gravity}

In Ref.\cite{Odintsov-lnR}, Nojiri and Odintsov proposed using a
single $\ln R$ term to explain both the current acceleration and
inflation. The Palatini formulation of $\ln R$ gravity is studied
in Ref.\cite{wang-lnR}.

The Modified Friedmann equation reads,
\begin{equation}
H^2=\frac{\kappa\rho_m+2\kappa\rho_r-\beta(\frac{R}{\beta}-\ln\frac{R}{-\alpha})}
{(1-\frac{\beta}{R})(6+3F(x)(1+\frac{1}{2}F(x)))} \label{d1}
\end{equation}
where $x\equiv\frac{\kappa\rho_m}{\beta}$ and $F$ is defined as
\begin{equation}
F(x)=\frac{3x}{(R(x)/\beta)^2-2R(x)/\beta}\ .\label{d2}
\end{equation}

It can be seen from equations (\ref{d1}) and (\ref{d2}) that when
$\beta\rightarrow 0$, the MF equation will reduce continuously to
the standard Friedmann equation. Thus, the $\ln R$ modification is
a smooth and continuous modification.

Let us first discuss the cosmological evolution without matter and
radiation. Define the parameter $n$ as $R_0=-\alpha e^{-n}$.
Substitute this to the vacuum field equation $L(R)=0$, we can get
$\alpha=e^n(2n+1)\beta$ and $R_0=-(2n+1)\beta$. Substitute those
to the vacuum MF equation and set $t=0$, we have
\begin{equation}
H_0^2=\frac{\beta(n+1)}{6(1+\frac{1}{2n+1})}\label{}
\end{equation}
Thus when $\beta\sim H_0^2\sim(10^{-33}eV)^2$ and $n>-1/2$, the
$\ln R$ modified gravity can indeed drive a current exponential
acceleration compatible with the observation. The role of the
parameter $\beta$ is similar to a cosmological constant or the
coefficient of the $1/R$ term in the $1/R$ gravity \cite{Wang}.

When the energy density of dust can not be neglected, i.e.
$\kappa\rho_m/\beta\gg 1$, $F(x)\sim0$ and if $\alpha$ satisfies
$|\ln(\kappa\rho_m/\alpha)|\ll\kappa\rho_m/\beta$, i.e.
$\exp(-\kappa\rho_m/\beta)\ll\alpha/\beta\ll\exp(\kappa\rho_m/\beta)$,
then $R\sim\ - \kappa\rho_m$. Then the MF equation (\ref{d1})
reduces to the standard Friedmann equation
\begin{equation}
H^2=\frac{\kappa}{3}(\rho_m+\rho_r)\label{}
\end{equation}
Thus if
$\exp(-\kappa\rho_{m,BBN}/\beta)\ll\alpha/\beta\ll\exp(\kappa\rho_{m,BBN}/\beta)$,
where $\rho_{m,BBN}$ is the energy density of dust in the epoch of
BBN, the $\ln R$ gravity can be consistent with the BBN
constraints on the form of Friedmann equation \cite{Carroll4}. One
possible choice is $\alpha=\beta$, for which the vacuum solution
can be solved exactly $R_0=-\alpha$. Since $\beta\sim H_0^2$, the
condition $\kappa\rho_m/\beta\gg 1$ breaks down only in recent
cosmological time. Thus the universe evolves in the standard way
until recently, when $\ln R$ term begins to dominate and drives
the observed cosmic acceleration.

\section{Conclusions and Discussions}
In this paper we reviewed the Palatini formulation of $L(R)$
gravity and its applications to cosmology. The nature of dark
energy is so mysterious that it is promising to seek further the
idea that there does not exist such mysterious dark energy, but it
is the General Relativity that fails to some extent at large scale.

The current "standard theory" of gravitation, Einstein's General
Relativity (GR) has  passed many ground-based and space satellite tests  within the Solar system experiments.
Any reasonable extended gravity models should consistently reduce to it at least in the weak field approximation.
We have derived the gravitational potential for the Palatini formulation of the modified gravity of the general L(R)
type,  which admits a de Sitter vacuum solution as current observations require, and conclude the the Newtonian
limit is always obtained in these class of models as well as that the deviations from GR is very small for a slowly
moving gravitational source \cite{mw}. Moreover, the running precision GR tests, like the Gravity Probe B and
the gravitational wave experiments Ligo/Lisa, will
present us more constraints to any extended gravity models, probably soon.
 However, to
reconcile the successful GR predictions within the solar system,
the extended gravity theories may be required to be scale
sensitive. It could be challenging and profound to locate the
additional curvature terms in our above discussions what form of the
scale dependence is.

\section*{Acknowledgements}
We would like to express our deep gratitude to Professor Sergei D.
Odintsov for numerous invaluable suggestions and comments when the
work we described in this paper is performed. We would also like
to thank Eanna E. Flanagan, Shin'ichi Nojiri, Xiang Ping Wu and Liu Zhao for helpful
discussions on those topics. This work is
supported partly by an ICSC-World Laboratory Scholarship, a China NSF and Doctoral
Foundation of National Education Ministry.

\begin{appendix}
\end{appendix}


\begin{thebibliography}{99}
\bibitem{Perlmutter} S. Perlmutter el al. Nature \textbf{404} (2000) 955;
Astroph. J. \textbf{517} (1999) 565; A. Riess et al. Astroph. J.
\textbf{116} (1998) 1009; ibid. \textbf{560} (2001) 49; Y. Wang,
Astroph. J. \textbf{536} (2000) 531.
\bibitem{Spergel} D.N.Spergel, et al., Astrophys.J.Suppl. \textbf{148} (2003) 1 [astro-ph/0302207]
; L.Page et al. astro-ph/0302220; M.Nolta, et al,
astro-ph/0305097; C.Bennett, et al, Astrophys.J.Suppl.
\textbf{148} (2003) 175 [astro-ph/0302209].
\bibitem{hui} B.Roukema, et al., A \& A \textbf{382} (2002)397: A.Lidz, et al., astro-ph/0309204 and references therein;
\bibitem{Peebles} P. J. E. Peebles, B. Ratra, astro-ph/0207347; S.M.Carroll, Living Rev. Rel.
\textbf{4} (2001) 1; T. Padmanabhan, Phys.
Rept. \textbf{380} (2003) 235.
\bibitem{Lue} A. Lue, R. Scoccimarro and G. Starkman,
astro-ph/0307034.
\bibitem{Dvali} G. Dvali, G. Gabadadze and M. Porrati, Phys. Lett.
B \textbf{485} (2000) 208.
\bibitem{Carroll} S. M. Carroll, V.Duvvuri, M.Trodden and M.
Turner, astro-ph/0306438; S. Capozziello, S. Carloni and A.
Troisi, "Recent Research Developments in Astronomy \&
Astrophysics" -RSP/AA/21-2003 [astro-ph/0303041]; S. Capozziello,
Int. J. Mod. Phys. D \textbf{11} (2002) 48.
\bibitem{Odintsov-string}
S. Nojiri and S. D. Odintsov, Phys. Lett. B \textbf{576} (2003)5.
\bibitem{Chiba}
T.Chiba, Phys. Lett. B \textbf{575} (2003) 1 [astro-ph/0307338].
\bibitem{Dick}
R.Dick, gr-qc/0307052; K.Freese and M.Lewis, Phys.Lett. B540 (2002) l; X.H.Meng and P.Wang, Comm.Theor.Phys.(2004)
at press, [hep-th/0310038]
\bibitem{Dolgov}
A. D. Dolgov and M. Kawasaki, Phys. Lett. B \textbf{573} (2003)1.
\bibitem{Vollick}
D. N. Vollick, Phys. Rev. D \textbf{68} (2003) 063510.
\bibitem{Wang}
X. H. Meng and P. Wang, Class. Quant. Grav. \textbf{20} (2003)
4949 [astro-ph/0307354]; ibid, Class. Quant. Grav. {\bf 21} (2004)
951 [astro-ph/0308031].
\bibitem{odintsov-insta}
S. Nojiri and S.D. Odintsov, Mod. Phys. Lett. A {\bf 19} (2004)
627.
\bibitem{new}
G. Allemandi, A. Borowiec and M. Francaviglia, hep-th/0403264; G.
J. Olmo and W. Komp, gr-qc/0403092; G. M. Kremer and D. S. M.
Alves, gr-qc/0404082.
\bibitem{Flanagan}
\'{E}.\'{E}.Flanagan, Phys. Rev. Lett. {\bf 92} (2004) 071101
[astro-ph/0308111]; ibid, Class. Quant. Grav. {\bf 21} (2003) 417;
ibid, gr-qc/0403063.
\bibitem{Odintsov-R2}
S. Nojiri and S. D. Odintsov, Phys. Rev. D \textbf{68} (2003) 123512
[hep-th/0307288].
\bibitem{Odintsov-lnR}
S. Nojiri and S. D. Odintsov, hep-th/0308176.
\bibitem{Wang2}
X. H. Meng and P. Wang, astro-ph/0308284.
\bibitem{Krauss}
L. M. Krauss and B. Chaboyer, astro-ph/0111597.
\bibitem{Fabris}
J. C. Fabris, S. V. B. Goncalves and P. E. de Souza,
astro-ph/0207430.
\bibitem{Linder}
E. V. Linder and A. Jenkins, astro-ph/0305286.
\bibitem{Melchiorri}
A. Melchiorri, L. Mersini, C.J. Odmann and M. Trodden,
astro-ph/0211522.
\bibitem{Sta}
A. A. Starobinsky, Phys.Lett.B {\bf 91} (1980) 99.
\bibitem{Odintsov-book}
I. L. Buchbinder, S. D. Odintsov and I. L. Shapiro, Effective
Action in Quantum Gravity, IOP Publishing, 1992.
\bibitem{Carroll4}
S. M. Carroll and M. Kaplinghat, Phys. Rev. D \textbf{65} (2002)
063507.
\bibitem{wang-lnR}
X. H. Meng and P. Wang, Phys. Lett. B {\bf 584} (2004) 1
[hep-th/0309062].
\bibitem{mw}
X. H. Meng and P. Wang, Gen. Rel. Grav.  {\bf 36} (2004) at press
[gr-qc/0311019]

\end{thebibliography}
\end{document}